\documentclass[a4paper,11pt]{article}

\usepackage{jinstpub}

\title{Towards portable muography with small-area, gas-tight glass Resistive Plate Chambers}

 \author[a,1]{S. Basnet, \note{Corresponding author}}
 \author[a]{E. Cortina Gil,}
 \author[a]{P. Demin,}  
 \author[a]{R.M.I.D. Gamage,}
 \author[a]{A. Giammanco,}
 \author[a]{M. Moussawi,}
 \author[b]{M. Tytgat,}
 \author[a]{S. Wuyckens.}
 
 \affiliation[a]{Centre for Cosmology, Particle Physics and Phenomenology (CP3),\\Universit\'e catholique de Louvain,\\Louvain-la-neuve, Belgium}
 
 \affiliation[b]{Department of Physics and Astronomy,\\Ghent University,\\Ghent, Belgium}

\emailAdd{samip.basnet@uclouvain.be}

\abstract{Imaging techniques that use atmospheric muons, collectively named under the neologism ``muography”, have seen a tremendous growth in recent times, mainly due to their diverse range of applications. The most well-known ones include but are not limited to: volcanology, archaeology, civil engineering, nuclear reactor monitoring, nuclear waste characterization, underground mapping, etc. These methods are based on the attenuation or deviation of muons to image large and/or dense objects where conventional techniques cannot work or their use becomes challenging.

In this context, we have constructed a muography telescope based on ``mini glass-RPC” planes following a design similar to the glass-RPC detectors developed by the CALICE Collaboration and used by the TOMUVOL experiment in the context of volcano radiography, but with smaller active area (16~$\times$~16 cm$^{2}$). The compact size makes it an attractive choice with respect to other detectors previously employed for imaging on similar scales. An important innovation in this design is that the detectors are sealed. This makes the detector more portable and solves the usual safety and logistic issues for gas detectors operated underground and/or inside small rooms. This paper provides an overview on our guiding principles, the detector development and our operational experiences. Drawing on the lessons learnt from the first prototype, we also discuss our future direction for an improved second prototype, focusing primarily on a recently adopted serigraphy technique for the resistive coating of the glass plates.  
}


\proceeding{XV Workshop on Resistive Plate Chambers and Related
Detectors - RPC2020 \\10-14 February 2020 \\Rome, Italy}

\begin{document}
\maketitle
\flushbottom

\section{Introduction}

First used as a probe to explore the inner part of solid structures in 1955 \cite{ancient}, cosmic muons, which are produced abundantly when primary cosmic rays interacts with the Earth's atmosphere, have been employed as an imaging tool in the subsequent decades. A variety of muon tomography (i.e., muography) methods have also been developed, exploiting new detection techniques and numerical algorithms for reconstruction and imaging. These methods are based on the attenuation or deviation of cosmic muons to image large and/or dense objects where conventional imaging techniques cannot work or their use becomes challenging. The steep rise of muography in recent times can be attributed, for the most part, to its range of applications across many different fields. These disciplines include but are not limited to: volcanology, archaeology, civil engineering, nuclear waste monitoring, underground mapping etc \cite{review}.  

In the last decade, several large muon telescopes with cross-sectional areas of $\mathcal{O}({1~m^{2}})$ have been developed mainly for applications in volcanology \cite{giulio-carlo,tanaka}, where the size of the object of interest requires the detectors to be at large distance. However, muography also finds applications in the imaging of smaller targets, such as nuclear waste monitoring and archaeology \cite{echia, rome}. For the latter case, the point of observation closest to the target is typically in narrow and confined environments (e.g., tunnels, underground chambers and crevasses etc.) with no power supply. This imposes various limitations on the system such as size of the detector, total weight including electronics, portability, and robustness. Moreover, it also provides some logistical challenges which include complete autonomy and low power consumption. Other teams have already been active in developing portable, robust, and autonomous detectors for the same use cases based on scintillating bars or fibers, nuclear emulsion, as well as gaseous detectors \cite{review}. Each detector type actually comes with its own set of benefits and drawbacks. Plastic scintillators are the most popular ones where spatial, and consequently angular resolution are not crucial. With the development of Silicon Photo Multiplier (SiPM) in recent years, robust, scintillator based muon trackers, suitable for harsh and logistically challenging environments, can now be designed at a much lower cost and power budget. However, a major drawback of SiPMs is a notable temperature dependence of the breakdown voltage (20-30 mV/$^{\circ}$C) that eventually affects gain, dark rate counts, and reverse current. With no power consumption, a limited cost, and excellent spatial $\mathcal{O}({1~\mu m})$ and angular $\mathcal{O}(1~mrad)$ resolutions, nuclear emulsion films, one of the earliest particle detectors, are perfect devices for certain muography applications. However, emulsion films cannot be triggered and start recording particle tracks right away from inception, and are also affected by variation in temperature and humidity. Another major drawback is the very specific equipment needed for the offline development and analysis of the emulsion plates \cite{review}.

Various gaseous detectors such as drift chambers \cite{drift}, micromegas \cite{micro}, resistive plate chambers (RPCs) \cite{tomuvol}, etc. have already been used in muography with some of these technical solutions being more popular for certain specific use cases. Here, a lower cosmic-ray flux is particularly advantageous since high resolution can easily be achieved using fewer, relatively simple electronic channels. However, there are some issues regarding the operation of such detectors in applications that are not under supervised laboratory condition. Since these detectors are normally operated with a continuous gas flow, gas bottles need to be replenished at the measurement site. Moreover, in some cases gas mixtures are used including quenchers and flammable components, posing significant safety issues in all underground applications. In particular large-area RPCs, while promising in many ways, have some specific drawbacks: they require a continuous flow of flammable gases, extremely high voltages ($\sim$ 10 kV) and a stable environmental operating conditions.With these issues in mind, we started developing a prototype of a portable muon telescope based on small and gas-tight glass RPCs (gRPCs), focusing mainly on small targets in contexts of challenging logistics.

\section{First Prototype and its Operational Experiences}

A first muon telescope prototype was developed in 2017-18 based on the design of the glass-RPC based Semi-Digital Hadronic Calorimeter developed within the CALICE collaboration \cite{calice} and the TOMUVOL project for muography \cite{tomuvol}. In contrast to the large-area gRPCs required by high-energy physics and volcanology cases, our first prototype is constructed with a much smaller active area of 16 $\times$ 16 cm$^{2}$. In order to make the detector more portable and also to solve the safety and logistic challenge affecting most gaseous detectors operated underground and/or inside small rooms, a specific guiding principle for our design is gas tightness. In addition, we also aim at ease of transportation of the entire set-up including the electronics. More detailed explanations on the experimental set-up, data collection, analysis and our experience while operating the prototype are provided in the following sections and in Ref. \cite{sophie}. 

\subsection{Experimental Setup}
\begin{figure}[htbp]
\centering
\includegraphics[width=0.85\textwidth]{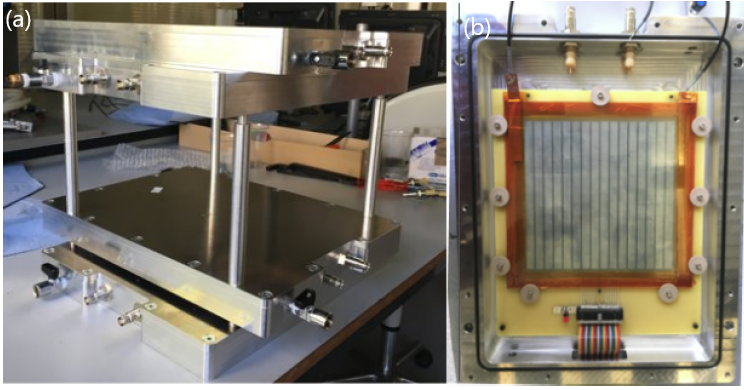}
\caption{\label{fig:i} (a) Our muon detector consisting of 4 mini-gRPC detector layers. \label{fig:ib} (b) One of the mini-gRPC detectors inside its casing.}
\end{figure}

Our first prototype consists of four layers of mini-gRPC detectors, as shown in the left panel of Figure \ref{fig:i}. Four removable spacer bars were used to create a distance of 14.8 cm between second and third plane for better track reconstruction. Each detector layer weighs 6.5 kg, contains 16 sensitive strips, with 1 cm pitch and 0.9 cm width, and has gas-gap of 1.1 mm, as can be seen in the right panel of Figure \ref{fig:i}. To achieve bidimensional information, the first and third detectors are aligned orthogonally to the second and the fourth. The gas mixture for operating the detector consists of R134a Freon (95.2\%), SF$_{6}$ (0.3\%) and isobutane (4.5\%) with a pressure slightly higher than the atmospheric one. 

Glass sheets with an area of 20~$\times$~20 cm$^{2}$ and thickness of 1.1 mm were used as electrodes. The outer side of each glass electrode was coated using a paint roller, with a colloidal dispersion of antimony-doped tin oxide in water (20\% powder and 80\% water) mixed with an equal quantity of methanol. To maintain a uniform gas gap, nine round edge spacers made up of polyether ether ketone (PEEK) were used between the glass sheets. The front-end electronic boards (FEB) used for this prototype are spare boards of the current Muon RPC system of the CMS experiment \cite{FEB1, FEB2}. Each FEB has four front-end modules with eight channels each (i.e., one FEB for two detector layers). Each channel consists of an amplifier with a charge sensitivity of 2 mV/fC, a discriminator, a monostable and a LVDS driver. The LVDS outputs of all the FEBs are connected to a System-on-Chip (SoC) module, which is installed on a carrier board with a wireless connection. 

To verify the gas-tightness of the chambers, tests were performed with a vacuum pump, measuring the rate of leakage after creating vacuum inside the chamber. Simultaneously, helium injection was used to localize potential leaking points and the necessary adjustments were made (e.g. the PCB connector had to be soldered as it was identified as a leaking point). In the end, a leak rate as low as 10$^{-9}$ mbar~l~s$^{-1}$ was achieved, compared to 10$^{-4}$ mbar~l~s$^{-1}$ before.

\subsection{Data Collection, Analysis and Results}

The working point of the detector was determined by performing two different scans in high voltage and threshold. From those scans, the optimal working point was determined to be at a high voltage of 6.6 -- 6.7 kV and a threshold of 105 -- 110 (arbitrary units). More details on the procedure are provided in Ref.~\cite{sophie}. Since the first and last strip in each detector were found to be very noisy, they were excluded from further analysis. Furthermore, in order to gather the coincidence data, the muon telescope was operated in a self-triggering mode, by imposing a logical AND of the four layers. 

\begin{figure}[htbp]
(a)
\includegraphics[width=.48\textwidth]{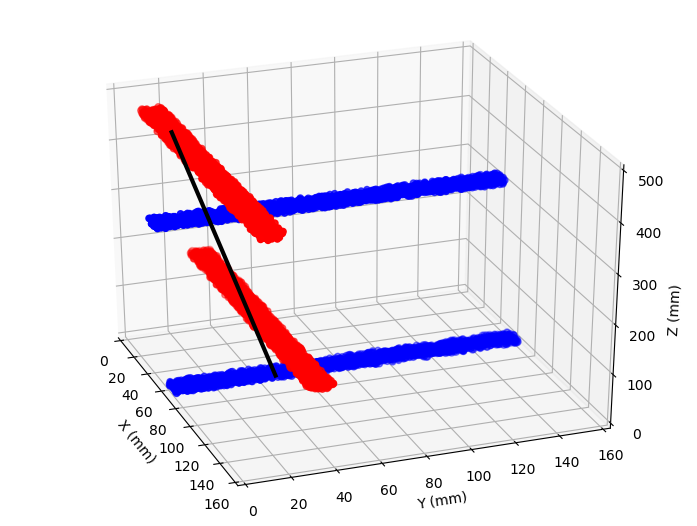}
(b)
\includegraphics[width=.49\textwidth]{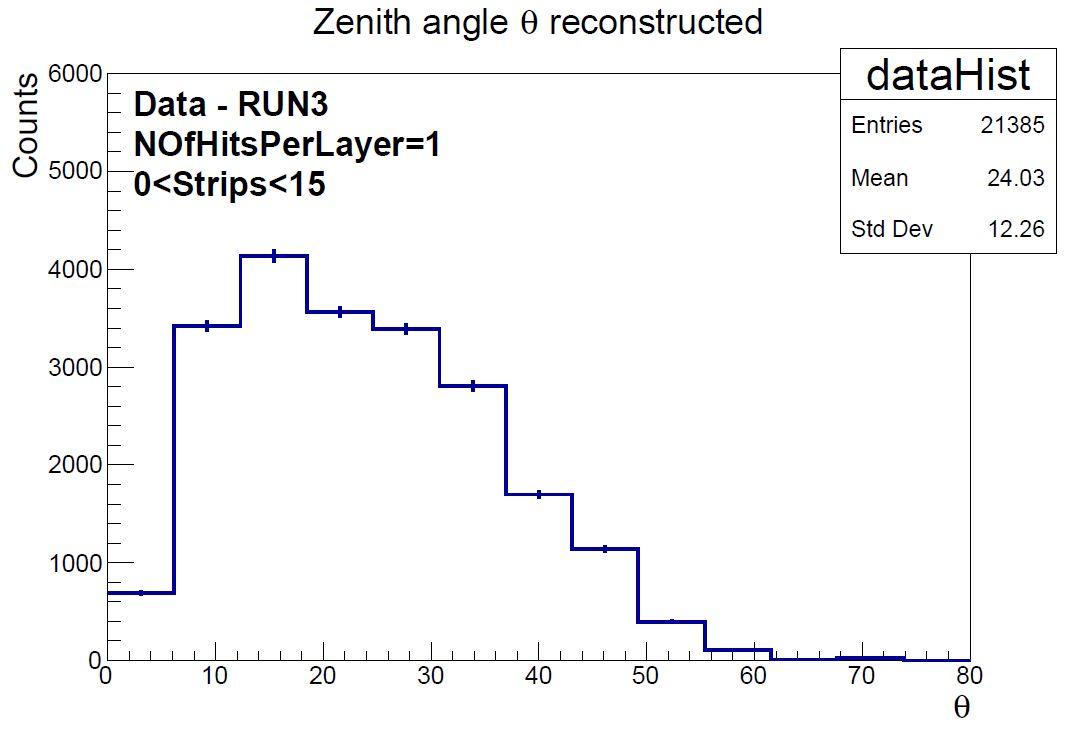}
\caption{\label{fig:ii} (a) Typical event showing the fired strips in each detector layer. X, Y, and Z represent spatial directions in mm. The black segment represents the reconstructed trajectory of a muon giving signal in the fired strips. (b) Distribution of reconstructed zenith angle collecting data for 6 days. }
\end{figure}

In the left panel of Figure \ref{fig:ii}, an illustrative event is shown with the reconstructed trajectory of a muon giving signal in all four planes. 
The distribution of the zenith angle reconstructed in events with exactly one hit per layer with hits present in each of the four layers after 6 days of data collection is shown in the right panel of Figure \ref{fig:ii}.

In an earlier stage of our project, when only two out of four detectors were ready, we had an opportunity to test several logistical aspects of our project in challenging operational outdoor conditions by participating in a two-week long life-on-Mars simulation at the Mars Desert Research Station (MDRS) in Utah, USA in March, 2018 \cite{mars}. The data collected from this campaign were suffering from issues (documented in~\cite{sophie}) that were solved only after return of the detector to our laboratory. However, we were able to verify the gas-tightness of the system as the pressure in the detector casing did not decrease significantly ($< 0.05$ atm) over a month. Also, the two detectors and the electronics (with total weight of 37.5 kg) did a successful round trip from Louvain-la-Neuve (Belgium) to a remote location in the Utah desert, validating its portability as well as robustness. 

\section{Current Work and Future Direction}
\subsection{Performance Studies}
Even though the experience with this first prototype has been encouraging, there are still various performance tests that need to be carried out with the current detector before the final design for an improved second prototype can be agreed upon. Previously, the data was taken without an external trigger, which made the efficiency and performance studies with the detector implausible and without having a good handle on the efficiency, pinning down the spatial resolution of our detector also proved difficult. In order to perform a reliable study of both the general performance and efficiency of our mini-gRPC detector, we have now added an external trigger system made up of two scintillating slabs with dimensions matching the active area of our gRPCs, placed above and below our telescope. Further work is also being done to include data from this external trigger in our current trigger and data acquisition framework. Once all these above-mentioned issues are sufficiently addressed, we plan to use our current detector to acquire muography data of a known structure, such as the CP3 building in Louvain-la-neuve. 

\subsection{Resistive Layer Coating Procedure}
While the work on the first prototype is still ongoing, preparations for the second prototype are in their initial phase. For our first prototype, the resistive coating was applied on the glass plates by hand using paint rollers. This manual procedure can introduce some undesirable non-uniformity in the coating quality (layer thickness, resistivity value) across the glass surface, which is known to adversely affect the performance of gRPCs. For our next prototype, a technique based on serigraphy is considered for the resistive layer coating of the glass plates.

\begin{figure}[htbp]
(a)
\includegraphics[width=.48\textwidth]{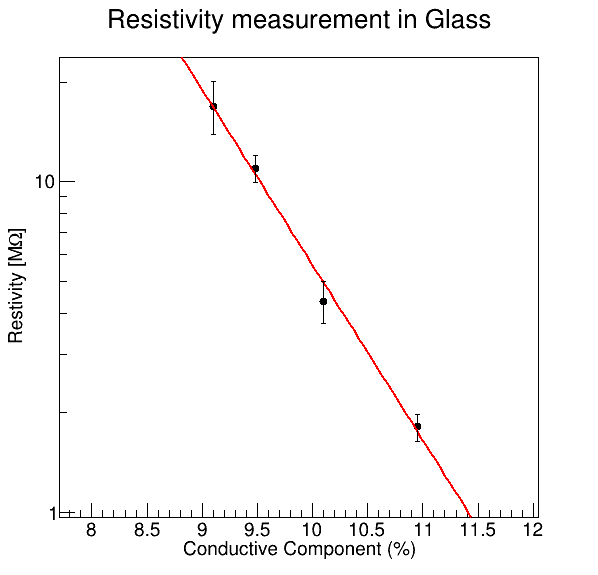}
(b)
\includegraphics[width=.49\textwidth]{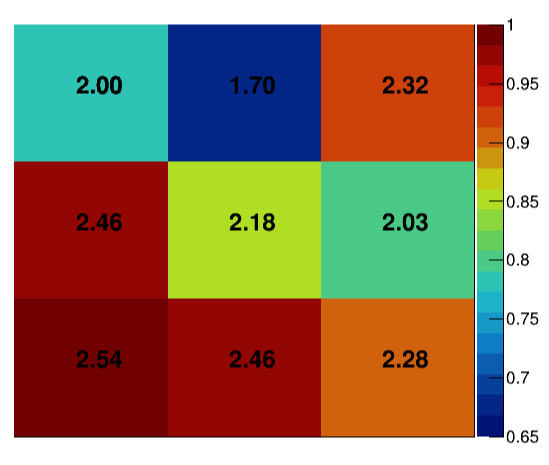}
\caption{\label{fig:iii} (a) Surface resistivity (in M$\Omega$) as a function of conductive component of the paste
(in \%). The black data points are used for fitting, resulting in the solid red line with equation $R = 13.94 \cdot e^{-1.22}$. (b) Surface resistivity (in M$\Omega$) measurement for one of the glass plates at 9 different locations, normalized to the highest reading.}
\end{figure}

Serigraphy, also called screen-printing, is a printing technique where a mesh is used to transfer paint into a surface, in our case glass plates, except in the areas that are made impermeable to the paint by blocking the stencil. Using the CEA facility in Saclay, France, we recently procured glass plates with uniform resistive layers, to be tested for our next prototype. To apply coating with serigraphy, the first step is to make the conductive compound, i.e. the paste, for which a mixture of conductive EDAG PM 404 and the EDAG 6017SS neutral component was used similar to the CALICE SDHCAL gRPCs \cite{calice_grpc}. The exact amount of each component in the mixture is determined by varying the amount of conductive component in the mixture and measuring the subsequent surface resistivity, which is shown along with an exponential fit in the left panel of Figure \ref{fig:iii}. For a surface resistivity of 2 M$\Omega$, $\sim$10.5\% conductive component was used in the mixture for our glass plates. Once the mixture is made, serigraphy is performed first on a kapton sheet and then on the glass surface, after which the glass is cured in the oven for 4-5 hours at 180$^{\circ}$C. The measurement of surface resistivity was done at nine different spots in the glass; a sample measurement is shown in the right panel of Figure \ref{fig:iii}. 

Next to serigraphy, a more common paint spraying technique will also be studied. Before taking a final decision on the method to be used for the construction of the second prototype, resistivity measurements of the coated glass plates will be performed over a longer time period to ensure sufficient stability.   

\subsection{Detector Readout}
Furthermore, in order to perform a high resolution muography, our next prototype aims at reaching a spatial resolution of $\mathcal{O}({1~mm})$ and timing resolution of $\mathcal{O}({1~ns})$, both of which are considered easily reachable for RPCs. This will require an increase in the number of readout channels, which in our setup is limited by the capability of our currently used FEBs. To this end, the use of the MAROC ASIC \cite{MAROC} is currently being studied. At the same time, the number of readout channels and the corresponding associated power consumption can be kept down via multiplexing, which allows to read a large number of strips with as few channels as possible by exploiting redundancy to compress the information in an unambiguous way. An example of such method that may be appropriate for our setup is known as genetic multiplexing \cite{multiplex}. Furthermore, alternative detector layouts are being studied to decrease the gas volume, and thus also the gas consumption, by using sealed chambers that include two resistive plates as outer layers.  

\section{Conclusion}

A first prototype of a telescope for muography applications to be operated in confined space was developed based on mini glass-RPC detectors. The system is intended to be low cost and portable, not only in terms of size and weight but also with respect to gas tightness and ease of transportation of the full setup, including electronics. The experience gained has been encouraging, as we were able to go from design to operation within a short amount of time. While the performance studies with the current detector are on-going, a new prototype will also be constructed soon -- drawing on the lessons learnt in the process -- with improvements in both the detector layout and fabrication as well as in the readout electronics.

\acknowledgments

This work was partially supported by the EU Horizon 2020 Research and Innovation Programme under the Marie Sklodowska-Curie Grant Agreement No. 822185, and by the Fonds de la Recherche Scientifique - FNRS under Grant No. T.0099.19. The corresponding author also acknowledges additional research grants from the FNRS - FRIA. We thank Dr. Stephan Aune and his serigraphy team at CEA, Saclay for helping us with resitive coating. 
We also thank the electronics group at the Center for Cosmology, Particle Physics, and Phenomenology (CP3), Universit\'{e} catholique de Louvain, for their help in setting up the external scintillator trigger.

\end{document}